# Mutual information between geomagnetic indices and the solar wind as seen by WIND – implications for propagation time estimates

T. K. March,[1] S. C. Chapman[1] and R. O. Dendy[2,1]

[1]Space and Astrophysics Group, Department of Physics, Warwick University, Coventry CV4 7AL, United Kingdom

[2]UKAEA Culham Division, Culham Science Centre, Abingdon, Oxfordshire OX14 3DB, United Kingdom.

### Abstract

The determination of delay times of solar wind conditions at the sunward libration point to effects on Earth is investigated using mutual information. This measures the amount of information shared between two timeseries. We consider the mutual information content of solar wind observations, from WIND, and the geomagnetic indices. The success of five commonly used schemes for estimating interplanetary propagation times is examined. Propagation assuming a fixed plane normal at 45 degrees to the GSE $x$-axis (i.e. the Parker Spiral estimate) is found to give optimal mutual information. The mutual information depends on the point in space chosen as the target for the propagation estimate, and we find that it is maximized by choosing a point in the nightside rather than dayside magnetosphere. In addition, we employ recurrence plot analysis to visualize contributions to the mutual information, this suggests that it appears on timescales of hours rather than minutes.

# 1  Introduction

An important problem in space weather forecasting and magnetospheric physics is the determination of solar wind effects on Earth, given measurements from a satellite. In particular we consider a spacecraft in orbit around the sunward libration point L1. Solar wind plasma passing this point takes around 60 minutes to reach the Earth, providing sufficient advance notice of solar wind conditions to facilitate real-time space weather prediction.

Early work on this subject [15, 8] demonstrated that an L1 monitor is a good predictor of solar wind conditions near Earth, assuming a planar description of the solar wind propagation. Studies of techniques to estimate plane orientation typically fall into two categories: 1) propagation of bulk properties [3, 12, 2, 18]; 2) propagation of specific features such as discontinuities [13] or sudden southward turnings of the magnetic field [7]. In case 1, correlation coefficients are computed between time-lagged solar wind timeseries from spatially separated spacecraft, in order to quantify the success of a particular time-lagging method. In case 2, a specific class of feature is identified in the timeseries from each spacecraft, and the difference between predicted and observed arrival times at the downstream spacecraft quantifies the degree of success.

In the present paper we adopt the first approach, considering also the related problem of what happens once the solar wind has propagated through interplanetary space and is incident on the Earth's magnetosphere. We take solar wind measurements from the WIND spacecraft [9, 11], but in contrast to previous work we look for correlation with detectable geomagnetic effects observed via the AE geomagnetic indices [5] on Earth. Estimation of propagation delays in interplanetary space has been well studied (see above). However, from an operational perspective the correlation between effects on Earth and data from an L1 monitor is also of direct interest, and is the focus of this paper. The method is not restricted by the requirement that a second spacecraft is in the solar wind during the same time period as the first, and hence we can expect to find more data available and thus obtain better statistics. Using L1 and Earth also provides a large distance over which to estimate the propagation, and hence greater contrast between the propagation estimates given by different methods.

# 2  Data

For the Earth's response to the solar wind we use provisional AE index data from 1995, around solar minimum. We use WIND data for the first half of 1995, from day 24 to day 200. During this period the spacecraft was in the vicinity of the L1 point. One advantage of using geomagnetic indices instead of a second spacecraft is that we are able to use a long period of data and are hence certain of obtaining a representative sample of the solar wind. We take magnetic field measurements from the MFI experiment [9], with 60s sampling, and proton bulk velocities and density from the SWE experiment [11] with around 98s sampling.

# 3  Method

Typically, a measurement from a spacecraft is used to specify the solar wind over a plane, whose normal $\mathbf{n}$ must be estimated. Let $\mathbf{v}$ denote the solar wind velocity, and $\mathbf{P}_W$ and $\mathbf{P}_E$ the spacecraft and Earth positions. The time for the solar wind to reach Earth is then given by [18]

$$\Delta t = \frac{(\mathbf{P}_W - \mathbf{P}_E).\mathbf{n}}{\mathbf{v}.\mathbf{n}} \qquad (1)$$

where we use the GSE (Geocentric Solar Ecliptic) coordinate system. In two-satellite studies, $\mathbf{P}_E$ would be the position of the target satellite; in the present work we use the Earth's position, but note that in practice it might be more appropriate to choose another target such as the northern cusp. We examine this choice in section 4.

We use equation (1) to generate timeseries (with 60s sampling rate) of time-lags to Earth from the WIND spacecraft. Different hypotheses concerning solar wind propagation, described below, give rise to different time-lag timeseries. To produce a projected signal, each time-lag timeseries is added to the original time index data of the signals measured by WIND. A projected signal is thus defined on an uneven timebase:

$$t \to t + \Delta t + \Delta t' \qquad (2)$$

Here the additional variable $\Delta t'$ is added to allow for the finite response time of the magnetospheric system. Our results are plotted with respect to this. In reality, the optimal $\Delta t'$ would be expected to depend on the state of the magnetosphere which in turn depends on solar wind conditions.

We employ the following hypotheses concerning the plane normal vector $\mathbf{n}$: 1) $\mathbf{n}$ is oriented along the $x$-axis; 2) $\mathbf{n}$ is perpendicular to $\mathbf{B}$ and lies in the ecliptic plane; 3) $\mathbf{n}$ is parallel to the vector $(B_y B_z, B_x B_z, B_x B_y)$ [13]; 4) $\mathbf{n}$ is in the ecliptic plane and is fixed perpendicular to the average magnetic field orientation, at $45^o$ to the $x$-axis (the Parker spiral angle); 5) $\mathbf{n}$ is perpendicular to $\mathbf{B}$, obtained using the minimum variance analysis (MVA) of [18], note that this is not a standard MVA [17]. For the projected timeseries we use the following solar wind parameters: 1) $B_x$; 2) $B_y$; 3) $B_z$; 4) $B$; 5) $v_x B_z$; 6) $v_x$; 7) $\epsilon = vB^2 \sin^4 \tan^{-1}(|\frac{B_y}{B_z}|)$; 8) $\rho$; 9) $\rho v$; 10) $\rho v^2$; here $\rho$ is the proton number density and the data are taken in the GSM (Geocentric Solar Magnetic) coordinate system.

In studies of propagation between spacecraft, a correlation coefficient $r$ [1] provides a linear measure of the correlation between the observed and projected timeseries. We employ the mutual information [16], which is sensitive to both linear and nonlinear correlations. For two integer timeseries $\mathbf{a}$ and $\mathbf{b}$, the mutual information (in units of bits) is defined by

$$I(\mathbf{a}, \mathbf{b}) = H(\mathbf{a}) + H(\mathbf{b}) - H(\mathbf{a}, \mathbf{b}) \qquad (3)$$

where $m$ is the number of bins used in the discretization to integers, and the entropy, $H$, is defined by

$$H(\mathbf{a}) = -\sum_i^m p_i^a \log_2 p_i^a \qquad (4)$$

$$H(\mathbf{a}, \mathbf{b}) = -\sum_i^m \sum_j^m p_{ij}^{a,b} \log_2 p_{ij}^{a,b} \qquad (5)$$

Here $p_i^a$ is the measured probability of observing an element of $\mathbf{a}$ with the value $i$, and $p_{ij}^{a,b}$ is the joint probability of observing timeseries $\mathbf{a}$ in bin $i$ and timeseries $\mathbf{b}$ in bin $j$. If the entropy $H$ is considered to be the amount of information gained about a system via a single observation, then the mutual information $I(\mathbf{a}, \mathbf{b})$ is the information gained about $\mathbf{b}$, on observation of $\mathbf{a}$. In order to ensure good statistics, the partitioning into $m$ bins is performed such that each bin contains the same number of data points. For comparison, a typical good linear correlation of $r = 0.8$, found in bulk plasma propagation studies (e.g. [4]), corresponds to a two bin ($m = 2$) mutual information of around 0.3 bits. For a better correlation of $r = 0.9$, $I$ would be around 0.5 bits, while for $r = 0.6$, $I$ would be approximately 0.1 bits.

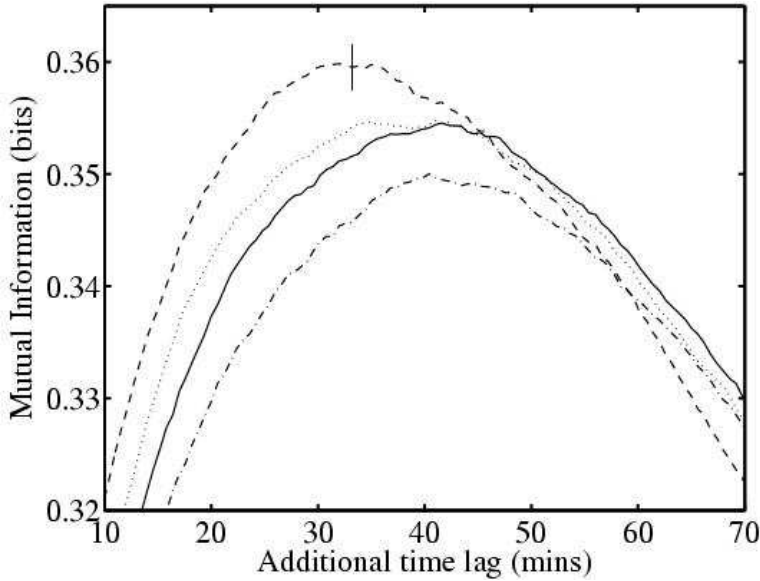

Figure 1: Eight bin mutual information between $v_x B_z$ and the AE index, plotted as a function of additional time-lag $\Delta t'$, for different time-lag methods, with WIND at L1. The solid line represents method 1, the $x$-axis method; dash-dot line: method 3, using the components of $\mathbf{B}$; dashed line: method 4, the Parker Spiral method. The dotted line represents methods 2 and 5, which lie so close together as to be indistinguishable on this scale. An error bar of 0.002 bits is shown.

## 4  Results

The mutual information is computed between the projected solar wind timeseries and the AE index, and expressed as a function of additional time-lag $\Delta t'$. After applying the time-lagging, the peak in mutual information as a function of $\Delta t'$ has a higher value, indicating improved correlation. Figure 1 shows a comparison between the time-lagging methods using eight bins. We consider the maximum mutual information obtained for each method, and conclude that the Parker Spiral method (method 4) in this case yields the best results with a peak in mutual information at $\Delta t' = 33 \pm 5$ minutes, with a peak value of $0.360 \pm 0.002$ bits, where the uncertainty is estimated following [14]. This value should be compared with the information contents (entropy) of the timeseries considered individually: 3 bits when using eight bins. We have also considered other constant plane angles but find that the Parker Spiral angle is optimal. The total magnetic field method (method 3) yielded the lowest mutual information. Interestingly, taking a different perspective, [13] found the opposite to be true when considering the error in predicting the arrival times of discontinuities. Of the ten different solar wind parameters considered, those which show a clear enhancement of the mutual information versus additional time-lag $\Delta t'$ are those involving $B_z$ and $v_x$, with the highest being $v_x B_z$; the peak mutual information between $\epsilon$ and AE is $0.0771 \pm 0.0007$ bits. We observe that the density $\rho$ does not show any significant correlation with the AE index.

Figure 2 shows the mutual information between the three geomagnetic timeseries AE, AU and AL and $v_x B_z$, results for the Dst index are also shown. The greatest mutual information is found for the AE index, defined as the difference between AU and AL. We note that the peak in mutual information for AU occurs at a smaller additional time-lag $\Delta t'$ than for AL. AU thus reacts earlier than AL, implying earlier transport of information into the dayside magnetosphere. Furthermore, the fact that AE responds more strongly than either AL or AU implies that AL and AU each posess different information about the solar wind driver.

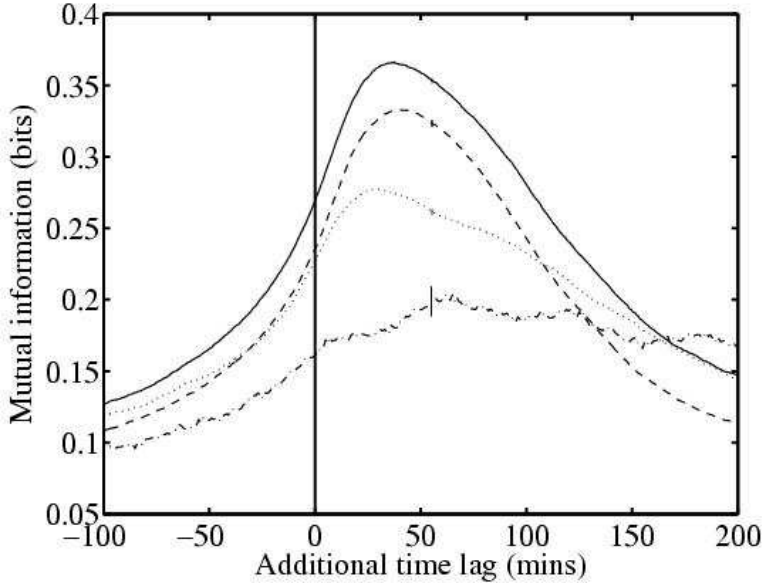

Figure 2: Eight bin mutual information, versus additional time-lag $\Delta t'$, between solar wind $v_x B_z$ and the geomagnetic indices AE (solid line), AU ( dotted line), AL ( dashed line) and also Dst (dash-dot line). Error bars are shown at $\Delta t' = 55$ mins. Note that Dst consists of hourly averaged data.

The construction of the time-lag timeseries requires a target position $\mathbf{P}_E$, see equation (1). For targeting Earth we use the origin of the GSE coordinate system. [18] use a position near the northern cusp, $(8, 0, 4)R_E$ where $R_E$ denotes one Earth radius. We assume for simplicity that once the solar wind impacts on the magnetosphere, the velocity information is lost and subsequent developments are best represented by a constant additional time-lag, although in reality we would expect this parameter to vary with magnetospheric conditions. Figure 3 shows the peak mutual information obtained between $v_x B_z$ and the AE index, using the $x$-axis propagation estimate, as a function of $\mathbf{P}_E(x)$. As the target is moved along the $x$-axis, a different additional time-lag $\Delta t'$ is required to maximize the mutual information. This maximum mutual information is not at a maximum around the northern cusp, however, in fact we observe stronger correlation than the cusp value in a region around 10 to 220 Earth radii downstream. Since we neglect the compression and deceleration of the solar wind as it impacts on the magnetosphere, our target position should be considered as the location of where the solar wind would have propagated to if the Earth had not been present. The peak in mutual information occurs with a target around 125 Earth radii downstream. Comparison with models of the dynamic magnetosphere might elucidate whether this is consistent with dayside reconnection being the dominant effect on Earth.

## 5 Visualization of Mutual Information

A useful method of visualizing nonlinear correlations between two timeseries is provided by recurrence plots [6], the statistics of which can be closely related to mutual information [10]. A recurrence plot is a 2D array $R_{ij}$ defined in terms of the $i$th and $j$th elements, $a_i$ and $a_j$, of a timeseries $\mathbf{a}$. For a discrete timeseries, where here we discretize into $m$ bins, we write

$$R_{ij} = |a_i - a_j| \qquad (6)$$

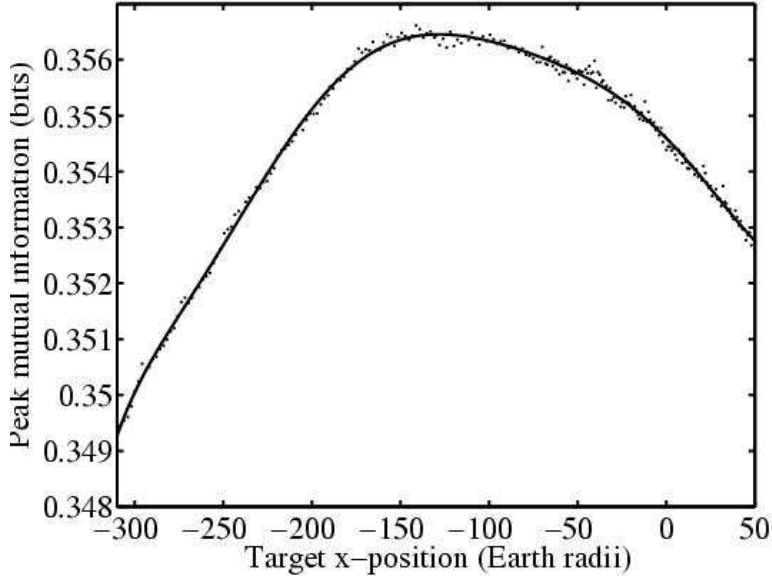

Figure 3: Peak mutual information between $v_x B_z$ and AE as a function of target position. Time lagging was performed using $x$-axis method. A polynomial fit is overlaid.

If two data points, $a_i$ and $a_j$, in the timeseries are in the same bin then the corresponding point in the array $R_{ij}$ takes the value 0 and is drawn as a black dot with values up to $m-1$ represented in shades of grey (or some chosen colourmap). A recurrence plot therefore displays which parts of a timeseries share a resemblance, with a resolution specified by $m$, and thus gives a graphical representation of any repetitive pattern in the timeseries.

Figure 4 shows colour recurrence plots for a short period (days 68 and 69 of 1995) of the AE index and $v_x B_z$ timeseries after time-lagging using the Parker spiral method. While there are many small scale differences between the plots, the overall large scale structure is similar. This shared pattern is quantified by the mutual information measurement, which can be related to counting the number of coordinates with black dots that are common to both plots [10]. Each black or white area corresponds to specific instants in time, so we can assess which parts of the timeseries contribute most strongly to the mutual information. In the present case it appears that the shared structure on the plots in Figure 4 derives from periods of both high and low activity, with structures which appear to be on a timescale of order hours.

# 6    Conclusions

We have quantitatively distinguished between hypotheses concerning time delay estimation using geomagnetic data. The mutual information quantifies the degree to which the far upstream solar wind shares information with, and hence presumably can be used to predict, AE and has also been used to estimate the response time of the magnetosphere to solar wind input. In this work we have assumed that the magnetospheric portion of the propagation can be represented by a constant time-lag. Since this assumption is employed for each propagation estimation method it should not affect the conclusions of the work.

Using this technique it is possible to examine the assumptions typically employed in time-lag estimation to Earth, which are inaccessible to multi-spacecraft studies. Here we have considered the target position and have shown that a tailward target seems to be optimal. Furthermore, using the technique of recurrence plot analysis, the nature of the correlation between the solar wind and geomagnetic timeseries as measured by mutual information can be visualized. This demonstrates that the correlation is dominated by phenomena on timescales of the order of a few hours, suggesting that substorms may be the dominant information carrier. An open question is whether the mutual information between AE and the solar wind input is strongly correlated with activity levels. Ordering the mutual information calculated over one day intervals with activity levels reveals a large amount of scatter which obscures any such trend. Elucidating this possible trend is thus a subset of future work. The mutual information technique on one interval alone cannot distinguish between a situation where there is a weak correlation present at all times or a situation where a strong correlation is present intermittently. However, the variability over day-long intervals and the common features on the recurrence plots (apparent on a variety of scales) imply the latter situation.

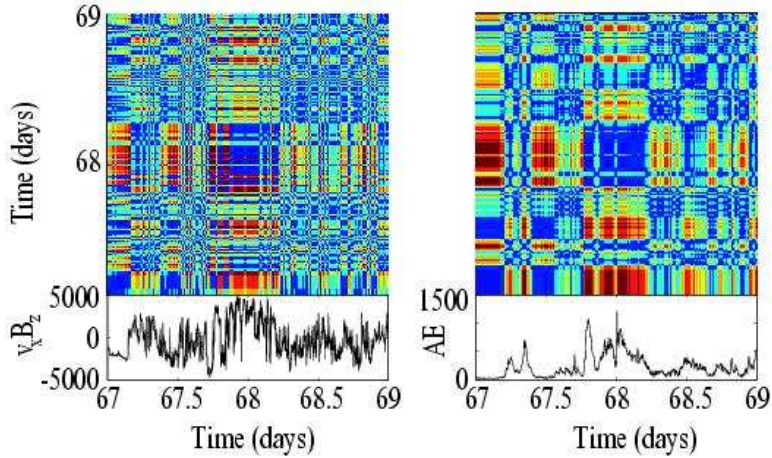

Figure 4: Recurrence plots of AE index and $v_x B_z$ for days 68 and 69 of 1995, after time-lagging $v_x B_z$ using the Parker Spiral method. Exact matches are coloured dark blue, with colours up to dark red representing increasing differences.

**Acknowledgements** The authors acknowledge support from the United Kingdom Engineering and Physical Sciences Research Council and the Particle Physics and Astronomy Research Council. We thank R P Lepping, A Szabo and K Ogilvie for provision of WIND MFI and SWE datasets, and the Kyoto WDC for provision of geomagnetic index data.